\documentclass[prepreint,twocolumn,trackchanges]{aastex63}
\hypersetup{linkcolor={blue},citecolor={blue}}
\usepackage{cprotect}

\received{June 1, 2019}
\revised{January 10, 2019}
\accepted{\today}
\submitjournal{ApJ}

\shorttitle{Light curve modeling of SN 2016aps}
\shortauthors{Suzuki et al.}

\begin{document}

\title{Extremely energetic supernova explosions embedded in a massive circumstellar medium: the case of SN 2016aps}

\correspondingauthor{Akihiro Suzuki}
\email{akihiro.suzuki@nao.ac.jp}

\author[0000-0002-7043-6112]{Akihiro Suzuki}
\affiliation{Division of Science, National Astronomical Observatory of Japan, 2-21-1 Osawa, Mitaka, Tokyo 181-8588, Japan}

\author[0000-0002-2555-3192]{Matt Nicholl}
\affiliation{Birmingham Institute for Gravitational Wave Astronomy, University of Birmingham, Birmingham, UK}
\affiliation{School of Physics and Astronomy, University of Birmingham, Birmingham, UK}
\affiliation{Institute for Astronomy, University of Edinburgh, Royal Observatory, Blackford Hill, UK}

\author[0000-0003-1169-1954]{Takashi J. Moriya}
\affil{Division of Science, National Astronomical Observatory of Japan, 2-21-1 Osawa, Mitaka, Tokyo 181-8588, Japan}
\affiliation{Center for Computational Astrophysics, National Astronomical Observatory of Japan, 2-21-1 Osawa, Mitaka, Tokyo 181-8588, Japan}
\affiliation{School of Physics and Astronomy, Faculty of Science, Monash University, Clayton, Victoria 3800, Australia}

\author[0000-0003-0304-9283]{Tomoya Takiwaki}
\affil{Division of Science, National Astronomical Observatory of Japan, 2-21-1 Osawa, Mitaka, Tokyo 181-8588, Japan}
\affiliation{Center for Computational Astrophysics, National Astronomical Observatory of Japan, 2-21-1 Osawa, Mitaka, Tokyo 181-8588, Japan}











\begin{abstract}

We perform one-dimensional radiation-hydrodynamic simulations of energetic supernova ejecta colliding with a massive circumstellar medium (CSM) aiming at explaining SN 2016aps, likely the brightest supernova observed to date. 
SN 2016aps was a superluminous Type-IIn SN, which released as much as $\gtrsim 5\times 10^{51}$ erg of thermal radiation. 
Our results suggest that the multi-band light curve of SN 2016aps is well explained by the collision of a $30\ M_\odot$ SN ejecta with the explosion energy of $10^{52}$ erg and a $\simeq 8\ M_\odot$ wind-like CSM with the outer radius of $10^{16}$ cm, i.e., a hypernova explosion embedded in a massive CSM.  
This finding indicates that very massive stars with initial masses larger than $40\ M_\odot$, which supposedly produce highly energetic SNe, occasionally eject their hydrogen-rich envelopes shortly before the core-collapse. We suggest that the pulsational pair-instability SNe may provide a natural explanation for the massive CSM and the energetic explosion. 
We also provide the relations between the peak luminosity, the radiated energy, and the rise time for interacting SNe with the kinetic energy of $10^{52}$ erg, which can be used for interpreting SN 2016aps-like objects in future surveys.

\end{abstract}



\section{Introduction} \label{sec:intro}
Core-collapse supernova (CCSN) explosions are one of the common luminous transient events in the optical sky. 
They are the explosive ejection of stellar mantle resulting from the iron core collapse of a massive star. 
The associated optical emission from the ejecta is normally powered by the initial shock-heating and/or the radioactive nickel freshly produced in the explosion. 
Since CCSNe are the final evolutionary stage of massive stars, the diversity of CCSNe reflects the variety of stellar properties, such as mass and metallicity, and activities, such as stellar winds and binary interaction, toward the final core-collapse. 
Transient surveys in the last few decades certainly revealed various modes of explosive stellar deaths including rare and bright SNe \citep{2011ApJ...743..114C,2011Natur.474..487Q,2012Sci...337..927G,2019ARA&A..57..305G}.

Recently, the superluminous Type-IIn SN 2016aps (also known as PS16aqy) was detected by the Panoramic Survey Telescope and Rapid Response System (Pan-STARRS; \citealt{2016arXiv161205560C}). 
\cite{2020NatAs.tmp...78N} reported its discovery and the results of the follow-up observations. 
SN 2016aps was classified as a Type-IIn SN (\citealt{1990MNRAS.244..269S,1997ARA&A..35..309F}; hereafter, SNe-IIn), which shows the spectral signature of the collision between SN ejecta and slowly-moving hydrogen-rich gas around the progenitor (circumstellar matter; CSM). 
At the redshift of $z=0.2657$, which is indicated by the hydrogen Balmer lines, the peak luminosity and the isotropic radiated energy of SN 2016aps reach $4.3\times 10^{44}$ erg s$^{-1}$ and $>5\times 10^{51}$ erg, making this object the brightest confirmed SN ever observed. 
The presence of the hydrogen line emission indicates SN 2016aps was likely powered by the CSM interaction. 
In the framework of the CSM-powered emission, the radiated energy of $5\times 10^{51}$ erg immediately indicates that the embedded SN should have been highly energetic, with an explosion energy larger than $5\times 10^{51}$ erg. 
Such highly energetic explosions are thought to occur in the so-called hypernovae, the exceptionally energetic explosions of very massive stars with the initial mass larger than $40\ M_\odot$ \citep{1998Natur.395..672I}. 
Other possibilities include even more massive stars ending up as the so-called pair-instability SNe \citep{1967PhRvL..18..379B,1967ApJ...148..803R} and SNe with an additional power source at its center \citep{2010ApJ...717..245K}. 

Based on the analytic scaling relations \citep{2011ApJ...729L...6C} and the single-zone light curve modeling by the Modular Open Source Fitter for Transients (\verb|MOSFiT|; \citealt{2018ApJS..236....6G}), \cite{2020NatAs.tmp...78N} argued that SN 2016aps was likely explained by the collision of $50$--$180\ M_\odot$ SN ejecta and $40$--$150\ M_\odot$ of CSM with the kinetic energy exceeding $10^{52}$ erg. 
The required total mass of the ejecta and the CSM exceeding $\sim 100\ M_\odot$ made authors suspect that SN 2016aps originated from a very massive star in the expected mass range of the pair-instability SNe ($\simeq 140$ -- $260\ M_\odot$; e.g., \citealt{2002ApJ...567..532H,2002ApJ...565..385U}). 
On the other hand, \cite{2015MNRAS.449.4304D} have conducted a series of multi-group radiation-hydrodynamic simulations of superluminous Type-IIn SNe and provided a couple of the models with the assumed kinetic energy of $10^{52}$ erg prior to the discovery of SN 2016aps. 
The model with the ejecta mass of $9.8\ M_\odot$ and the CSM mass of $17.3\ M_\odot$ roughly explains the peak luminosity and the evolutionary timescale of SN 2016aps, which is in contrast to the results of \cite{2020NatAs.tmp...78N} requiring much larger mass. 
 
For assessing the previous results and pining down the appropriate model parameters more precisely, a more wide and systematic model parameter survey based on radiation-hydrodynamic simulations is required. 
Recently, we systematically studied the CSM-powered SNe with a wide range of the model parameters; the ejecta mass and energy, and the CSM mass and radius \citep{2020ApJ...899...56S}. 
In the study, we provide the peak luminosity, the radiated energy, and the rise time of the CSM-powered emission and compared them with the observed samples of SNe-IIn \citep{2014ApJ...788..154O,2020A&A...637A..73N}. 
However, the assumed ejecta energy was up to $2\times 10^{51}$ erg and was not sufficient to explain the superluminous SN 2016aps. 
In this study, we perform the model parameter survey for highly energetic SN ejecta with the explosion energy of $10^{52}$ erg by using the same model setup as \cite{2020ApJ...899...56S}. 
We provide the relations between the peak bolometric luminosity, the radiated energy, and the rise time for the models explored, which are also beneficial for interpreting other SN 2016aps-like events detected in on-going and future transient surveys.

This paper is structured as follows. 
First, we consider how the quantities characterizing the system can roughly be constrained by analytical estimations in Section \ref{sec:parameter_estimate}. 
We then describe the numerical setups and the parameter sets explored in this study in Section \ref{sec:setup}. 
In Section \ref{sec:results}, we present the simulation results with a particular focus on the model best explaining SN 2016aps. 
The implications of the results are discussed in Section \ref{sec:discussion}. 
Finally, Section \ref{sec:summary} provides a summary of this study. 

\section{Parameter constraints by analytic estimation}\label{sec:parameter_estimate}
Before delving into the detailed model parameter survey with numerical simulations, we introduce the model parameters characterizing the system and how the parameter space can be narrowed down by analytic estimations. 
We note that similar discussion has been repeated in past studies for (semi-)analytic and numerical light curve modelings of SNe-IIn \citep[e.g.,][]{2011ApJ...729..143C,2011MNRAS.415..199M,2013MNRAS.428.1020M,2012ApJ...746..121C,2012ApJ...757..178G,2012ApJ...759..108S,2014ApJ...790L..16M,2014ApJ...789..104O,2019ApJ...884...87T,2020PASJ...72...67T}. 
As we shall describe in Section \ref{sec:setup}, our numerical simulations are based on \cite{2020ApJ...899...56S}, in which the important parameters characterizing the system are the ejecta mass $M_\mathrm{ej}$ and energy $E_\mathrm{sn}$, and the CSM mass $M_\mathrm{csm}$ and radius $R_\mathrm{csm}$. 

We first consider the energy budget of the thermal radiation. 
Since the isotropic-equivalent radiated energy of SN 2016aps is at least $\sim 5\times 10^{51}$ erg, the initial kinetic energy should be much larger than the normal explosion energy of $\sim 10^{51}$ erg. 
Such highly energetic explosions with the explosion energy of the order of $10^{52}$ erg are usually only expected for very massive stars with the initial mass more massive than 40 $M_\odot$ , i.e., hypernovae, which was introduced to explain Type-Ic SNe with broad line spectral features \citep{1998Natur.395..672I}. 
Therefore, we consider the explosions of such massive progenitors with the ejecta mass of the order of $10\ M_\odot$. 
Even though the initial kinetic energy is of the order of $10^{52}$ erg, producing as much as $5\times 10^{51}$ erg of the radiation energy requires a highly efficient conversion of the kinetic energy to thermal radiation.  
This requires the CSM mass to be at least a considerable fraction of the ejecta mass so that the CSM efficiently decelerates the expanding ejecta and liberates its kinetic energy. 

Next, the evolutionary timescale of SN 2016aps gives a characteristic length scale. 
For interaction-powered SNe, the rising timescale of a light curve reflects the photospheric radius. 
At the time of the shock breakout, the diffusion velocity of photons produced at the shock front is equal to the shock velocity $v_\mathrm{sh}$. 
Therefore, the timescale of photon diffusion from the shock front to the photosphere at a radius $r=R_\mathrm{ph}$ is approximately given by $t_\mathrm{dif}\simeq R_\mathrm{ph}/v_\mathrm{sh}$. 
The rising time of SN2016aps is $t_\mathrm{rise}\simeq 40$ days \citep{2020NatAs.tmp...78N}. 
Since the typical velocity of the ejecta is $(2E_\mathrm{sn}/M_\mathrm{ej})^{1/2}\simeq 10^9$ cm s$^{-1}$ for $E_\mathrm{sn}=10^{52}$ erg and $M_\mathrm{ej}=10\ M_\odot$, the requirement $t_\mathrm{dif}\simeq t_\mathrm{rise}$ leads to the following constraint on the photospheric radius,
\begin{equation}
    R_\mathrm{ph}\simeq 3\times 10^{15}
    \left(\frac{M_\mathrm{ej}}{10\ M_\odot}\right)^{-1/2}
    \left(\frac{E_\mathrm{sn}}{10^{52}\ \mathrm{erg}}\right)^{-1/2}\ \mathrm{cm}.
\end{equation}
This is roughly consistent with the almost constant photospheric radius, $\sim 5\times 10^{15}$ cm, estimated by the multi-band light curve of SN 2016aps at late epochs \citep{2020NatAs.tmp...78N}. 
For an infinitely extended CSM with the density profile given by $\rho=Ar^{-2}$, the photospheric radius of $R_\mathrm{ph}=5\times 10^{15}$ cm requires the coefficient $A$ to be
\begin{equation}
    A\simeq 10^{16}
    \left(\frac{R_\mathrm{ph}}{5\times 10^{15}\ \mathrm{cm}}\right)
    \left(\frac{\kappa}{0.34\ \mathrm{cm}^2\ \mathrm{g}^{-1}}\right)^{-1}
    \ \mathrm{g\ cm}^{-1},
\end{equation}
where a constant electron scattering opacity $\kappa=0.34$ cm$^2$ g$^{-1}$ is assumed. 
This value is much higher than those of normal stellar winds, $A=10^{11}$--$10^{12}$ g cm$^{-1}$ \citep{2014ARA&A..52..487S}. 
However, the mass of the optically thick CSM, $4\pi AR_\mathrm{ph}\simeq 0.5\ M_\odot$, is not large enough to realize a high kinetic-to-radiation conversion efficiency for an ejecta mass of the order of $10\ M_\odot$. 
Instead, this can be achieved with a higher CSM density in the inner regions, with a distribution that is truncated around the photospheric radius. 
Therefore, we expect that the massive CSM of the order of $10\ M_\odot$ is confined within a radius of several $10^{15}$--$10^{16}$ cm.

\section{Numerical methods}\label{sec:setup}
\subsection{Simulation setups}
We conduct 1D radiation-hydrodynamic simulations by using the same numerical code as \cite{2020ApJ...899...56S} (see also \citealt{2019ApJ...887..249S} for details). 
The numerical setups of our simulations are similar to \cite{2020ApJ...899...56S}, but with different input parameters. 

\subsubsection{Initial conditions}
The simulations are performed in the 1D spherical coordinate $r$. 
We consider a freely expanding spherical SN ejecta with a broken power-law density profile. 
The inner and outer density slopes are set to $d\ln\rho/d\ln r=-1$ and $-10$, respectively \citep{1989ApJ...341..867C,1999ApJ...510..379M}. 
The initial kinetic energy $E_\mathrm{sn}$ and the ejecta mass $M_\mathrm{ej}$ specify the density and velocity scales.  
The characteristic velocity separating the inner and outer parts of the ejecta is given as a function of the ejecta mass $M_\mathrm{ej}$ and the kinetic energy $E_\mathrm{sn}$ by
\begin{equation}
    v_\mathrm{br}=1.2\times 10^9
    \left(\frac{M_\mathrm{ej}}{10\ M_\odot}\right)^{-1/2}
    \left(\frac{E_\mathrm{sn}}{10^{52}\ \mathrm{erg}}\right)^{-1/2}
    \ \mathrm{cm}\ \mathrm{s}^{-1},
\end{equation}
for the adopted density structure. 
We start our simulations at $t=10^3$ s. 
Initially, we assume that the local internal energy density is $5\%$ of the local kinetic energy density. 
Therefore, the pressure of the ejecta does not affect the dynamical evolution as long as the ejecta expands in an adiabatic way. 
The ejecta initially extends to the inner radius of the CSM at $r=R_\mathrm{in}=4\times 10^{12}$ cm. 

We assume a wind-like CSM with an outer exponential cut-off,
\begin{equation}
    \rho_\mathrm{csm}(r)=
    \frac{pM_\mathrm{csm}}{4\pi \Gamma(1/p)R_\mathrm{csm}r^2}
    \exp\left[-\left(\frac{r}{R_\mathrm{csm}}\right)^p\right],
    \label{eq:rho_csm}
\end{equation}
where $p=10$ and $\Gamma(x)$ is a Gamma function. 
The CSM mass and radius, $M_\mathrm{csm}$ and $R_\mathrm{csm}$, thus specify the density structure. 
The CSM distribution has an exponential cut-off around $r=R_\mathrm{csm}$, beyond which we assume a normal steady wind, $\rho_\mathrm{out}(r)=A_\mathrm{out}r^{-2}$ with $A_\mathrm{out}=5\times 10^{11}$ g cm$^{-1}$, up to the outer boundary of the numerical domain at $r=R_\mathrm{out}=1.28\times 10^{17}$ cm. 

The interface between the inner dense CSM and the outer normal wind is one of the uncertainties associated with the modeling of interaction-powered SNe. 
The inner dense CSM may be produced by an extensive mass-loss with an ejection velocity faster than the preceding normal wind. 
In this case, the impact of the massive ejection produces the forward and reverse shocks, and the contact surface in between. 
On the other hand, the inner dense CSM may also be created by an enhanced mass-loss before the core-collapse while the wind velocity remains similar to that of the normal wind. 
In this case, the CSM-wind interface may be smoothly connected rather than the abrupt change in the density structure. 
Our current setting would rather correspond to the latter situation.

We separate the ejecta from the other components (CSM and normal wind) by introducing the ejecta mass fraction $X_\mathrm{ej}$, which is initially set to
\begin{equation}
    X_\mathrm{ej}=
    \left\{
    \begin{array}{ccl}
    1&\mathrm{for}&r\leq R_\mathrm{in}\\
    0&\mathrm{for}&R_\mathrm{in}<r.
    \end{array}\right.
\end{equation}
The spatial distribution of the ejecta mass fraction evolves as a passive scalar along with the density $\rho(t,v)$ and the radial velocity field $v(t,r)$,
\begin{equation}
    \frac{\partial (\rho X_\mathrm{ej})}{\partial t}
    +\frac{1}{r^2}\frac{\partial (r^2\rho v X_\mathrm{ej})}{\partial r}=0,
\end{equation}
in a standard manner. 

\subsubsection{Radiative processes\label{sec:radiative_process}}
The chemical composition of the ejecta and the ambient matter is characterized by the hydrogen and helium mass fractions of $X_\mathrm{h}$ and $X_\mathrm{he}$. 
Then, the following absorption and scattering coefficients,
\begin{equation}
    \kappa_\mathrm{a}=3.7\times 10^{22}\chi_\mathrm{ion}(1+X_\mathrm{h})(X_\mathrm{h}+X_\mathrm{he})\rho T_\mathrm{g}^{-7/2}\ \mathrm{cm^2\ g^{-1}},
    \label{eq:kappa_a}
\end{equation}
(the local density $\rho$ and the gas temperature $T_\mathrm{g}$ are in cgs units; see, e.g., \citealt{1979rpa..book.....R}),
and
\begin{equation}
    \kappa_\mathrm{s}=0.2(1+X_\mathrm{h})\chi_\mathrm{ion}\ \mathrm{cm}^2\ \mathrm{g}^{-1},
    \label{eq:kappa_es}
\end{equation}
corresponding to free-free absorption and electron scattering, are assumed. 
Here the factor $\chi_\mathrm{ion}$ represents the effect of hydrogen recombination and is given by
\begin{equation}
\chi_\mathrm{ion}=\frac{1}{1+(T_\mathrm{g}/T_\mathrm{rec})^{-\beta}},
\end{equation} 
with $\beta=11$ \citep{2019ApJ...879...20F}, which drastically reduces the opacity at $T_\mathrm{g}<T_\mathrm{rec}$. 

Most of the following simulations assume hydrogen-rich gas with $X_\mathrm{h}=0.73$ and $X_\mathrm{he}=0.25$, and the recombination temperature of $T_\mathrm{rec}=7\times 10^3$ K throughout the numerical domain. 
Although the observations of SN 2016aps suggest that the CSM is hydrogen-rich, the embedded SN ejecta may be hydrogen-poor. 
In order to clarify the effects of different chemical compositions and opacity in the ejecta, we try a simulation with the same free parameters as the best-fit model (see below), but with helium-rich composition in the ejecta. 
In the case of helium-rich ejecta, the hydrogen and helium mass fractions are assumed to be
\begin{equation}
    X_\mathrm{h}=0.73(1-X_\mathrm{ej}),
\end{equation}
and
\begin{equation}
    X_\mathrm{he}=0.25(1-X_\mathrm{ej})+0.98X_\mathrm{ej},
\end{equation}
which realizes helium-rich gas, $X_\mathrm{he}=0.98$, with the reduced absorption and scattering coefficients (Equations \ref{eq:kappa_a} and \ref{eq:kappa_es}) in the ejecta ($X_\mathrm{ej}\simeq 1$), while the ambient gas remains hydrogen-rich ($X_\mathrm{ej}\ll1$). 
We also consider a higher recombination temperature in the ejecta,
\begin{equation}
T_\mathrm{rec}=7\times 10^3(1-X_\mathrm{ej})+1.2\times 10^4X\mathrm{ej},
\end{equation}
(in K).  

We also consider the effects of other opacity sources. 
We only assume free-free absorption and electron scattering, and neglect bound-bound opacity. 
While this is a good approximation in the early evolution of ejecta, bound-bound transitions increasingly become an important opacity source as the ejecta cools. 
In the numerical light curve modeling of SNe, the effects of bound-bound transition are sometimes considered by introducing the so-called opacity floor \citep[e.g.,][]{1988A&A...196..141S,2011ApJ...729...61B,2015ApJ...814...63M}. 
We carry out a simulation with the same parameter as the best-fit model, but with a floor value of $\kappa_\mathrm{a,floor}=0.01$ cm$^2$ g$^{-1}$ \citep{2011ApJ...729...61B} added to the absorption coefficient (Equation \ref{eq:kappa_a}).

\subsubsection{Light curve calculations}
The bolometric light curve is directly obtained from the simulation. 
We use the time evolution of the outgoing radial flux $F_\mathrm{r}$ at $R_\mathrm{obs}=10^{17}$ cm to calculate the bolometric light curve,
\begin{equation}
    L_\mathrm{bol}(t)=4\pi R_\mathrm{obs}^2F_\mathrm{r}(t,R_\mathrm{obs}),
\end{equation}
in the source rest frame.

While we first explore appropriate parameter sets for SN 2016aps by bolometric light curve fitting, the multi-band light curves are available and provide us with information on the color evolution. 
Therefore, we also perform post-process ray-tracing calculations by using the simulation results to obtain the multi-band light curves. 
For some models of interest, we obtain the radial distributions of the hydrodynamic variables every $5\times 10^5$ s. 
We perform the ray-tracing calculation for each snapshot assuming that the ejecta is at rest while photons propagate. 
This approximation is justified when the maximum ejecta velocity is much smaller than the speed of light $c$. 
The details of the numerical procedures are found in Appendix \ref{sec:post_process}.

\subsection{Models}
\begin{table}
\begin{center}
  \caption{Model descriptions}
\begin{tabular}{lrrr}
\hline\hline
Series&$M_\mathrm{ej}[M_\odot]$&$E_\mathrm{sn}[10^{51}\mathrm{erg}]$&$R_\mathrm{csm}[10^{15}\mathrm{cm}]$
\\
\hline
\verb|M10E10R5|&$10.0$&$10.0$&$5.0$\\
\verb|M20E10R5|&$20.0$&$10.0$&$5.0$\\
\verb|M30E10R5|&$30.0$&$10.0$&$5.0$\\
\verb|M40E10R5|&$40.0$&$10.0$&$5.0$\\
\verb|M10E10R10|&$10.0$&$10.0$&$10.0$\\
\verb|M20E10R10|&$20.0$&$10.0$&$10.0$\\
\verb|M30E10R10|&$30.0$&$10.0$&$10.0$\\
\verb|M40E10R10|&$40.0$&$10.0$&$10.0$\\
\verb|M10E10R20|&$10.0$&$10.0$&$20.0$\\
\verb|M20E10R20|&$20.0$&$10.0$&$20.0$\\
\verb|M30E10R20|&$30.0$&$10.0$&$20.0$\\
\verb|M40E10R20|&$40.0$&$10.0$&$20.0$\\
\hline\hline
\end{tabular}
\label{table:model_description}
\end{center}
\end{table}
\begin{table}
\begin{center}
  \caption{Photospheric radii for fully ionized hydrogen-rich CSMs}
\begin{tabular}{lrrr}
\hline\hline
&\multicolumn{3}{c}{$R_\mathrm{ph}/R_\mathrm{csm}\ \mathrm{for}\ R_\mathrm{csm}/10^{15}\mathrm{cm}=$}
\\
$M_\mathrm{csm}[M_\odot]$&
$5$&
$10$&
$20$
\\
\hline
$1$&$0.664$&$0.354$&$0.123$\\
$2$&$0.783$&$0.514$&$0.218$\\
$3$&$0.838$&$0.605$&$0.293$\\
$4$&$0.871$&$0.664$&$0.354$\\
$5$&$0.894$&$0.706$&$0.404$\\
$6$&$0.911$&$0.738$&$0.446$\\
$7$&$0.924$&$0.763$&$0.482$\\
$8$&$0.935$&$0.783$&$0.514$\\
$9$&$0.944$&$0.800$&$0.541$\\
$10$&$0.952$&$0.815$&$0.564$\\
$12$&$0.965$&$0.838$&$0.605$\\
$14$&$0.975$&$0.856$&$0.637$\\
$16$&$0.984$&$0.871$&$0.664$\\
$18$&$0.991$&$0.883$&$0.687$\\
$20$&$0.997$&$0.894$&$0.706$\\
$22$&$1.00$&$0.903$&$0.723$\\
$24$&$1.01$&$0.911$&$0.738$\\
$26$&$1.01$&$0.918$&$0.751$\\
$28$&$1.02$&$0.924$&$0.763$\\
$30$&$1.02$&$0.930$&$0.774$\\
$40$&$1.03$&$0.952$&$0.815$\\
$50$&$1.04$&$0.968$&$0.843$\\
\hline\hline
\end{tabular}
\label{table:Rph}
\end{center}
\end{table}

We perform series of simulations with various parameter sets. 
As in \cite{2020ApJ...899...56S}, we regard simulations with the same $M_\mathrm{ej}$, $E_\mathrm{sn}$, and $R_\mathrm{csm}$, but with different $M_\mathrm{csm}$ as one model series. 
For each model series, the adopted mass grid is $1\ M_\odot\leq M_\mathrm{csm}\leq 10\ M_\odot$ by a step of $1\ M_\odot$, $10\ M_\odot\leq M_\mathrm{csm}\leq 30\ M_\odot$ by a step of $2\ M_\odot$, $M_\mathrm{csm}=40\ M_\odot$, and $M_\mathrm{csm}=50\ M_\odot$. 
In total, a single model series contains 22 individual models with different $M_\mathrm{csm}$. 

We fix the initial kinetic energy to be $E_\mathrm{sn}=10^{52}$ erg. 
As we have estimated in Section \ref{sec:parameter_estimate}, the CSM radius is several $10^{15}$--$10^{16}$ cm. 
Therefore, we examine the following three cases, $R_\mathrm{csm}=5\times 10^{15}$, $10^{16}$, and $2\times 10^{16}$ cm. 
For each CSM radius, we change the ejecta mass from $M_\mathrm{ej}=10\ M_\odot$ to $40\ M_\odot$ by a step of $10\ M_\odot$. 
Each model series is named after the adopted values of the parameters, which are listed in Table \ref{table:model_description}. 
Table \ref{table:Rph} presents the scattering photosphere divided by the CSM radius, $R_\mathrm{ph}/R_\mathrm{csm}$, calculated assuming fully ionized gas for each set of the CSM mass and radius.

\section{Results}\label{sec:results}
\subsection{Light curve properties}
\begin{figure}
\begin{center}
\includegraphics[scale=0.4]{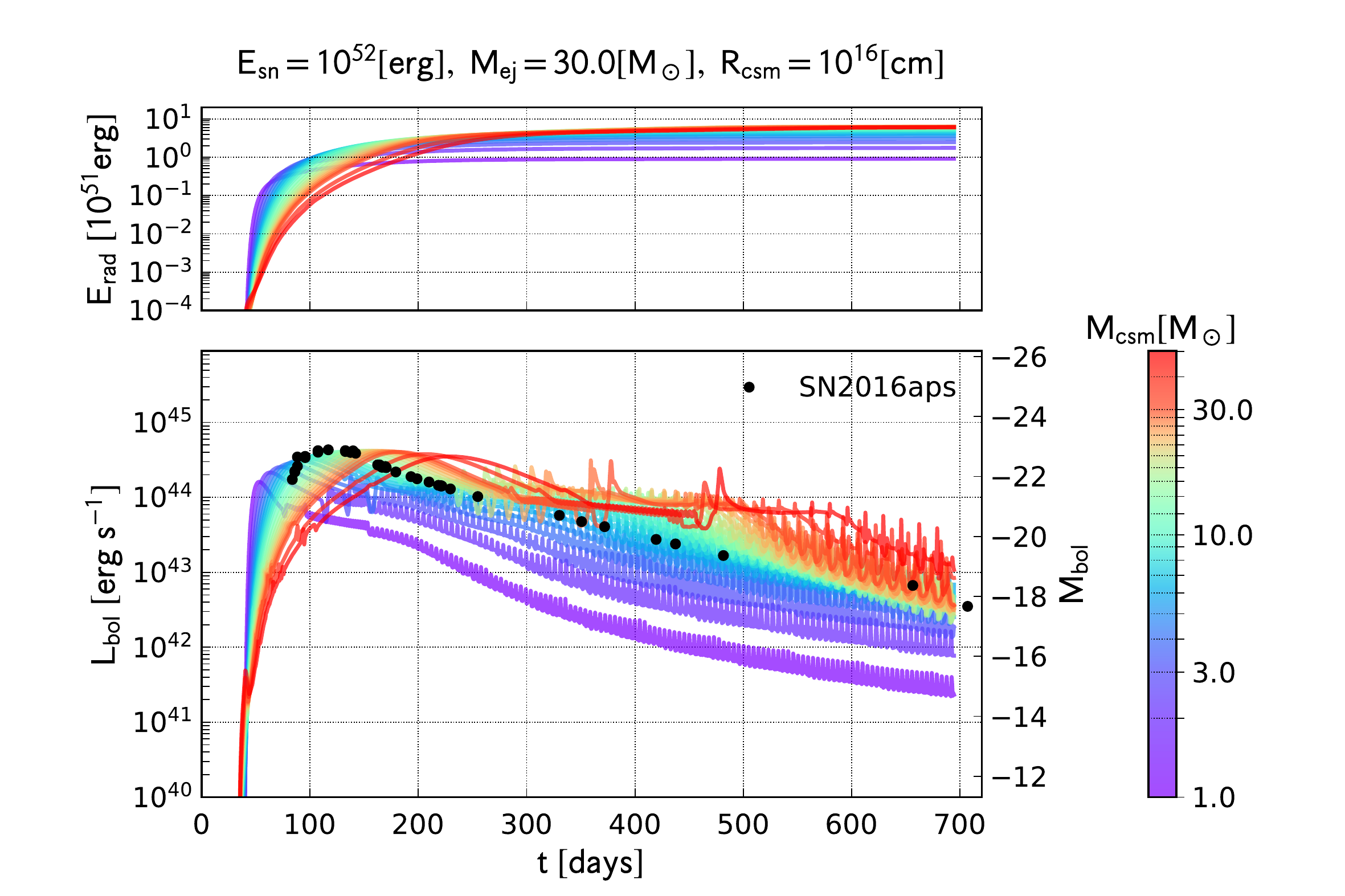}
\cprotect\caption{Cumulative radiated energy (upper panel) and bolometric light curves (lower panel) of the models in \verb|M30R10E10|. 
The model parameters are set to $E_\mathrm{sn}=10^{52}$ erg, $M_\mathrm{ej}=20\ M_\odot$, and $R_\mathrm{csm}=10^{16}$ cm. 
The bolometric light curve of SN 2016aps is also plotted in the lower panel. 
}
\label{fig:lc}
\end{center}
\end{figure}
\begin{figure*}
\begin{center}
\includegraphics[scale=0.48]{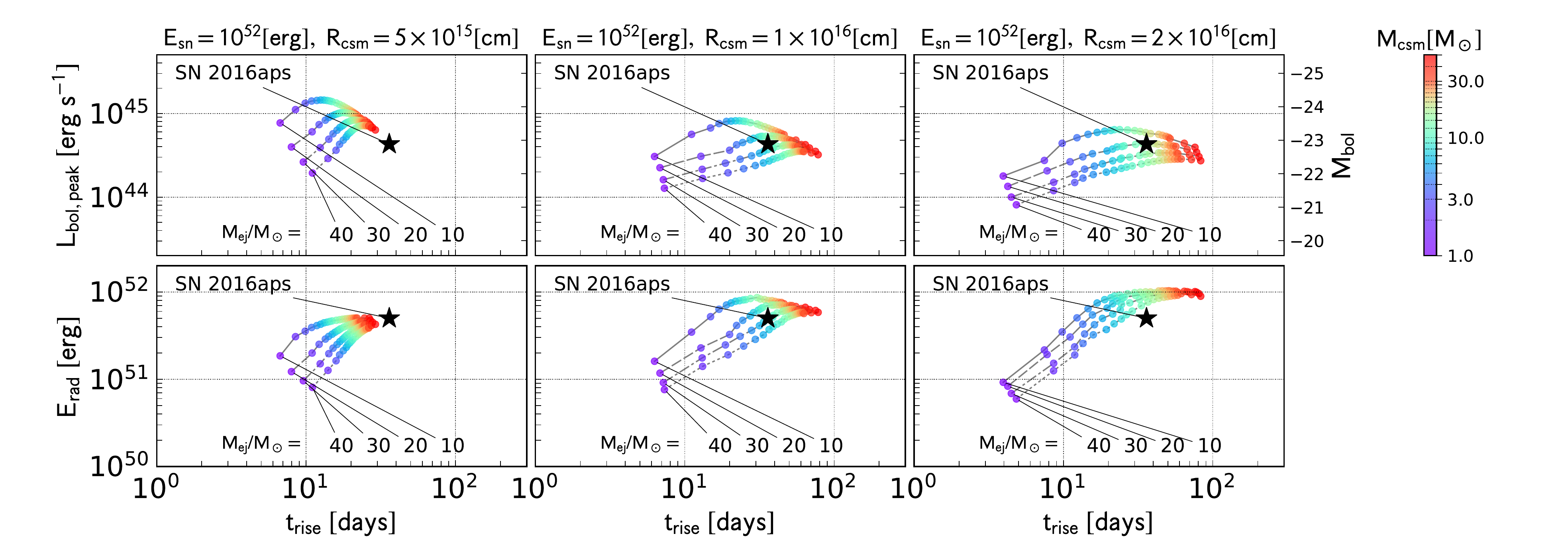}
\cprotect\caption{Peak luminosity $L_\mathrm{bol,peak}$ (upper panels) and radiated energy $E_\mathrm{rad}$ (lower panels) as a function of the rising time $t_\mathrm{rise}$. 
The results for all the model explored in this study is plotted. 
The left, center, and the right columns present the results for the models with $R_\mathrm{csm}=5\times 10^{15}$, $10^{16}$, and $2\times 10^{16}$ cm. 
}
\label{fig:Lpeak_Tpeak}
\end{center}
\end{figure*}

The bolometric light curves are obtained for all the models explored in this study. 
Figure \ref{fig:lc} shows example light curves for the model series \verb|M30R10E10|, compared with that of SN 2016aps \citep{2020NatAs.tmp...78N}. 
With these model parameters, the bolometric luminosity reaches its maximum value around several $10^{44}$ erg s$^{-1}$. 
The evolutionary timescale of the CSM-powered emission is predominantly determined by the photon diffusion timescale in the CSM. 
Therefore, it monotonically increases with CSM masses. 
For the CSM mass range explored, $1.0\leq M_\mathrm{csm}/M_\odot\leq 50$, the timescale varies in a wide range from several to $\sim100$ days. 
As seen in Figure \ref{fig:lc}, some models agree with the peak bolometric luminosity and the evolutionary timescale of SN 2016aps.

For a more quantitative and systematic comparison with SN 2016aps, we introduce some quantities characterizing the bolometric light curves as in \cite{2020ApJ...899...56S}. 
First, we define the peak bolometric luminosity $L_\mathrm{bol,peak}$ as the maximum value of the bolometric luminosity. 
The epoch of the peak luminosity is denoted by $t_\mathrm{peak}$. 
Next, we calculate the total radiated energy $E_\mathrm{rad}$ by integrating the bolometric light curve up to the end of the simulation, $t=6\times10^7$ s. 
Finally, we introduce the rise time $t_\mathrm{rise}$. 
We define the rise time $t_\mathrm{rise}$ so that the luminosity reaches a fraction $f_\mathrm{rise}$ of the peak value at $t_\mathrm{peak}-t_\mathrm{rise}$, $L_\mathrm{bol}(t_\mathrm{peak}-t_\mathrm{rise})=f_\mathrm{rise} L_\mathrm{bol,peak}$. 
For SN 2016aps, the bolometric luminosity already reaches $\sim40$\% of the peak value at the first detection, $\simeq36$ days before the peak (source rest-frame; \citealt{2020NatAs.tmp...78N}). 
Thus, we set the fraction to $f_\mathrm{rise}=0.4$. 

We calculate these characteristic quantities for all the models and show the resultant $L_\mathrm{bol,peak}$--$t_\mathrm{rise}$ and $E_\mathrm{rad}$--$t_\mathrm{rise}$ relations in Figure \ref{fig:Lpeak_Tpeak}. 
The behaviors of the relations for a single model series (models with different $M_\mathrm{csm}$) are similar to those examined by \cite{2020ApJ...899...56S}. 
The peak luminosity increases for shorter $t_\mathrm{rise}$ and decreases for longer $t_\mathrm{rise}$. 
These two parts with the different behaviors are separated by the condition $M_\mathrm{csm}\simeq M_\mathrm{ej}$. 
Models with larger $M_\mathrm{csm}$ dissipate a larger fraction of the kinetic energy and produce larger $E_\mathrm{rad}$. 
The $L_\mathrm{bol,peak}$--$t_\mathrm{rise}$ and $E_\mathrm{rad}$--$t_\mathrm{rise}$ relations for different model series have systematic offsets from each other. 
The general trend is that the peak luminosities are higher for smaller CSM radii $R_\mathrm{csm}$ and ejecta mass $M_\mathrm{ej}$. 
On the other hand, the rise time becomes shorter for smaller CSM radii $R_\mathrm{csm}$ and the ejecta mass $M_\mathrm{ej}$. 

These relations are compared with the rise time, the peak luminosity, and the radiated energy of SN 2016aps ; $t_\mathrm{rise}=36$ days, $L_\mathrm{bol,peak}=4.3\times 10^{44}$ erg s$^{-1}$, and $E_\mathrm{rad}=5\times 10^{51}$ erg. 
As seen in Figure \ref{fig:Lpeak_Tpeak}, the models with $R_\mathrm{csm}=5\times 10^{15}$ cm (left column) do not explain the rise time of SN 2016aps, while the models with $R_\mathrm{csm}=2\times 10^{16}$ cm (right column) overshoot the total radiated energy of SN 2016aps. 
Some models with $R_\mathrm{csm}=10^{16}$ cm (center column) satisfy the observational constraints. 
We find that the model with $M_\mathrm{csm}=8$ $M_\odot$ in the model series \verb|M30E10R10| best reproduces the light curve properties of SN 2016aps (referred to as the best-fit model, hereafter). 

\begin{figure}
\begin{center}
\includegraphics[scale=0.55]{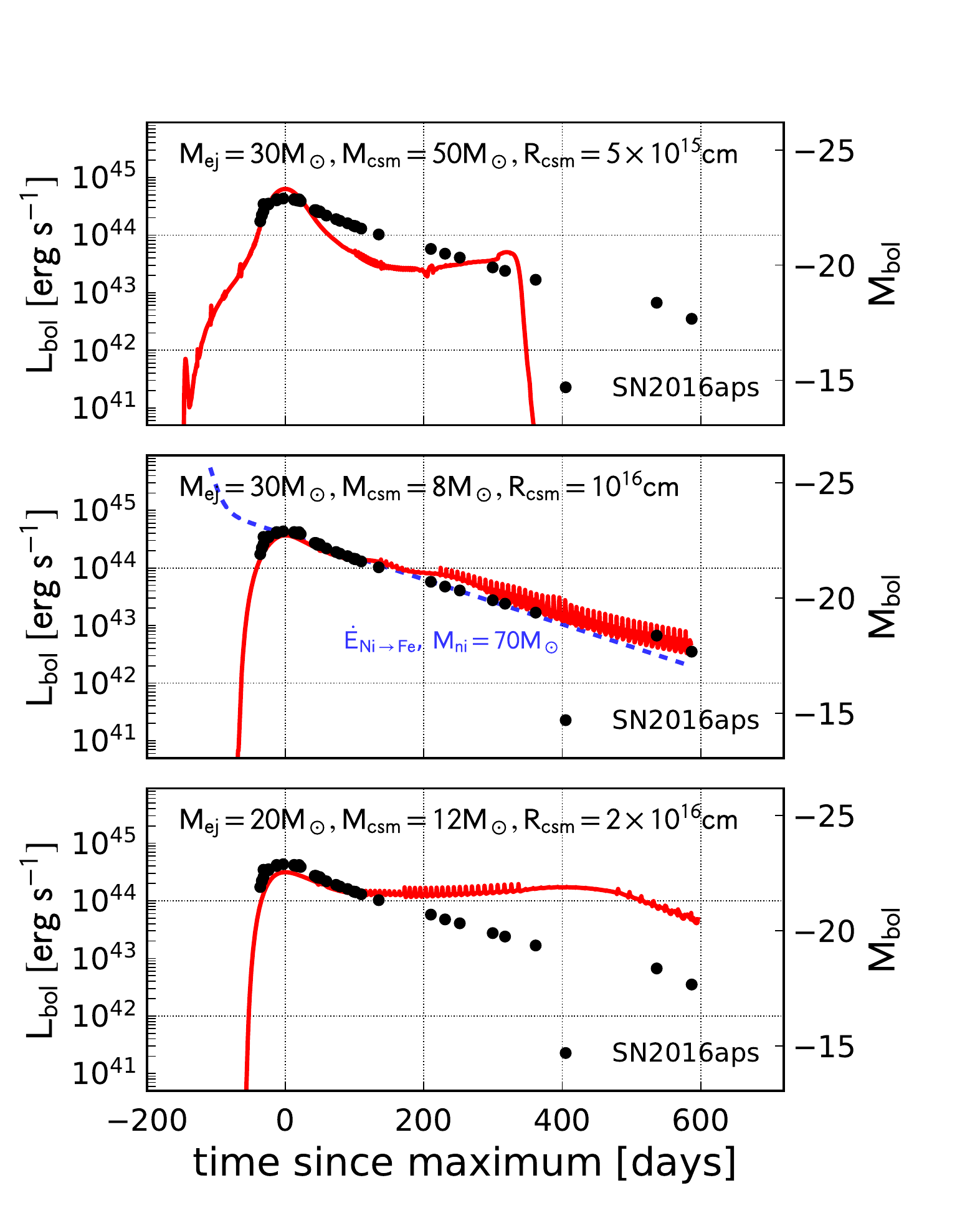}
\cprotect\caption{Bolometric light curves of some models with the peak bolometric luminosity and rise time similar to SN 2016aps. 
In the middle panel, the energy deposition rate for radioactive $^{56}$Ni with a nickel mass of $70\ M_\odot$ is plotted for comparison (blue dashed line).}
\label{fig:lc_comparison}
\end{center}
\end{figure}

We pick up some models with the peak luminosity and the rise time similar to those of SN 2016aps and compared them with the observed bolometric light curve in Figure \ref{fig:lc_comparison}. 
In the middle panel of Figure \ref{fig:lc_comparison}, we plot the bolometric light curve of the best-fit model. 
The model light curve around the peak successfully reproduces the observed light curve. 
The tail of the light curve suffers from numerical oscillations, which often happens when the ejecta becomes dilute and cool at late epochs. 
Nevertheless, the overall declining trend of the model light curve is similar to that of SN 2016aps. 
In the same panel, we also show the energy deposition rate of $^{56}$Ni radioactive decay \citep{1994ApJS...92..527N} for the purpose of comparison. 
Although the decline rate of the light curve around $200$ days matches the $^{56}$Co decay, it requires a huge nickel mass of $70\ M_\odot$. 
The top and bottom panels of Figure \ref{fig:lc_comparison} show the models with similar $L_\mathrm{bol,peak}$ and $t_\mathrm{rise}$ but with the smaller and larger $R_\mathrm{csm}$ as constrained from the diagram in Figure \ref{fig:Lpeak_Tpeak}.  
Although the two models explain the observed light curve around the peak, both of them exhibit clear deviations at late epochs. 
In the model with the smaller $R_\mathrm{csm}=5\times 10^{15}$ cm (top panel), the bolometric luminosity suddenly drops $\sim 350$ days after the peak. 
This is because of the truncated CSM at $R_\mathrm{csm}=5\times 10^{15}$ cm, beyond which the shock dissipation cannot produce thermal radiation efficiently. 
On the other hand, the late-time bolometric luminosity of the model with the larger $R_\mathrm{csm}=2\times 10^{16}$ cm significantly exceeds the observed luminosity. 
This is because the CSM is extended into a relatively large radius and the shock dissipation continuously produces thermal photons. 
This explains why the models with $R_\mathrm{csm}=2\times 10^{16}$ cm overshoot the radiated energy of SN 2016aps around $t_\mathrm{rise}=30$-$40$ days in Figure \ref{fig:Lpeak_Tpeak} (right panels). 
Therefore, we conclude that $R_\mathrm{csm}=10^{16}$ cm is appropriate. 

\begin{figure}
\begin{center}
\includegraphics[scale=0.55]{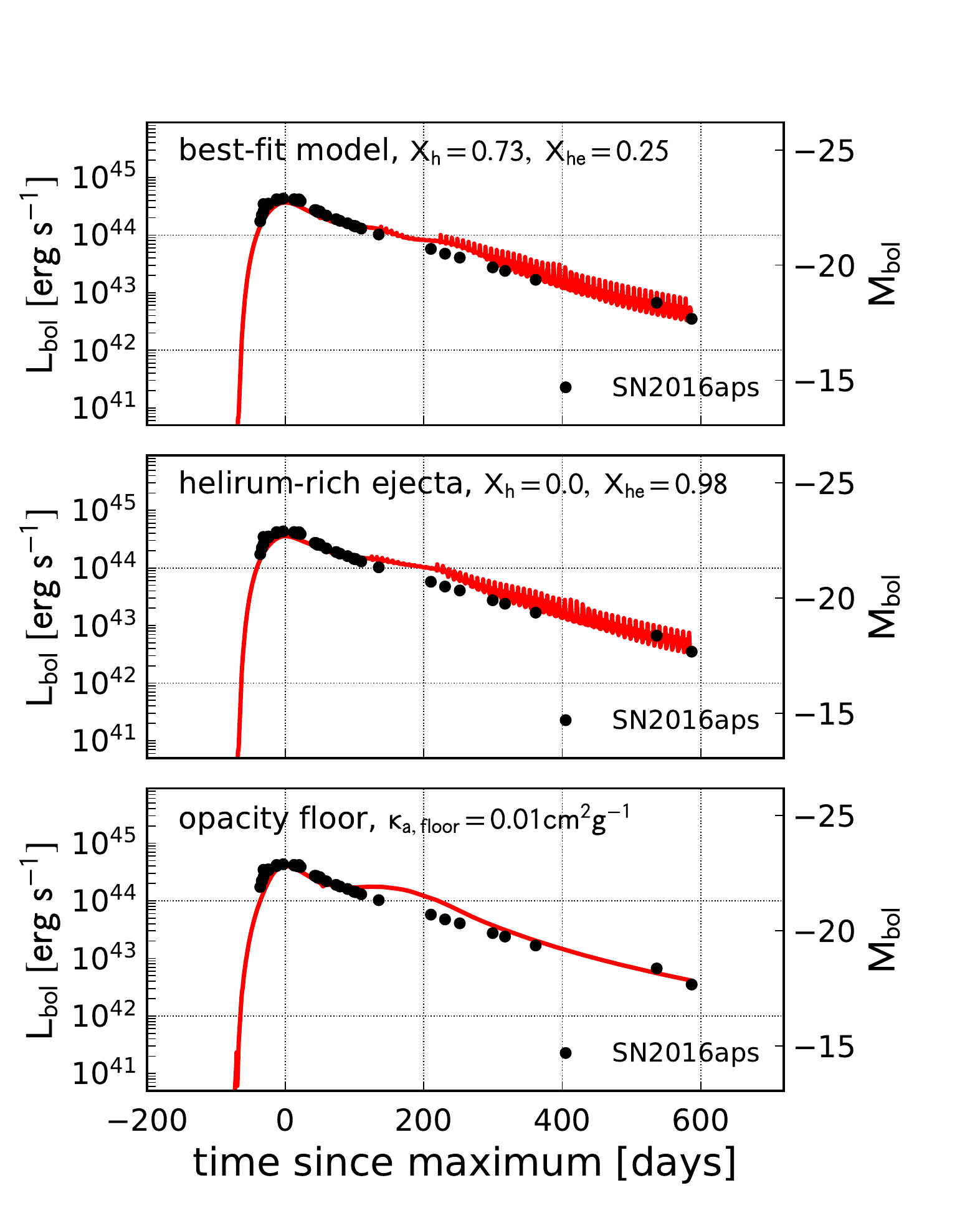}
\cprotect\caption{Bolometric light curves of the models with different opacities. 
In the top panel, the best-fit model (same as the middle panel of Figure \ref{fig:lc_comparison}) is plotted. 
In the middle and bottom panels, the model with the same parameters as the best-fit model, but with the helium-rich ejecta and the floor value for the absorption opacity are plotted, respectively. }
\label{fig:opacity_effect}
\end{center}
\end{figure}

\subsection{Opacity effects}
As we have described in Section \ref{sec:radiative_process}, we carry out a couple of simulations with the same parameters as the best-fit model, but with different values of the opacity. 
In one model, the helium-rich composition is assumed for the embedded ejecta. 
The other model assumes hydrogen-rich composition, but an opacity floor imitating bound-bound opacity is added to the absorption coefficient. 
The bolometric light curves of these two additional models are compared with that of the best-fit model in Figure \ref{fig:opacity_effect}. 
The models show a negligible difference around the peak. 
This is anticipated because both effects become important when the inner ejecta is revealed or hydrogen recombination reduced the free-free opacity down to the assumed floor value. 
The bolometric luminosity around $150$--$200$ days after the maximum shows deviation from the best-fit model. 
Nevertheless, the difference in the bolometric luminosity is not so significant and less likely to influence the estimate of the free parameters. 
Therefore, we focus on models with hydrogen-rich medium throughout the numerical domain in the following.

\subsection{Evolution of hydrodynamic variables}

\begin{figure}
\begin{center}
\includegraphics[scale=0.55]{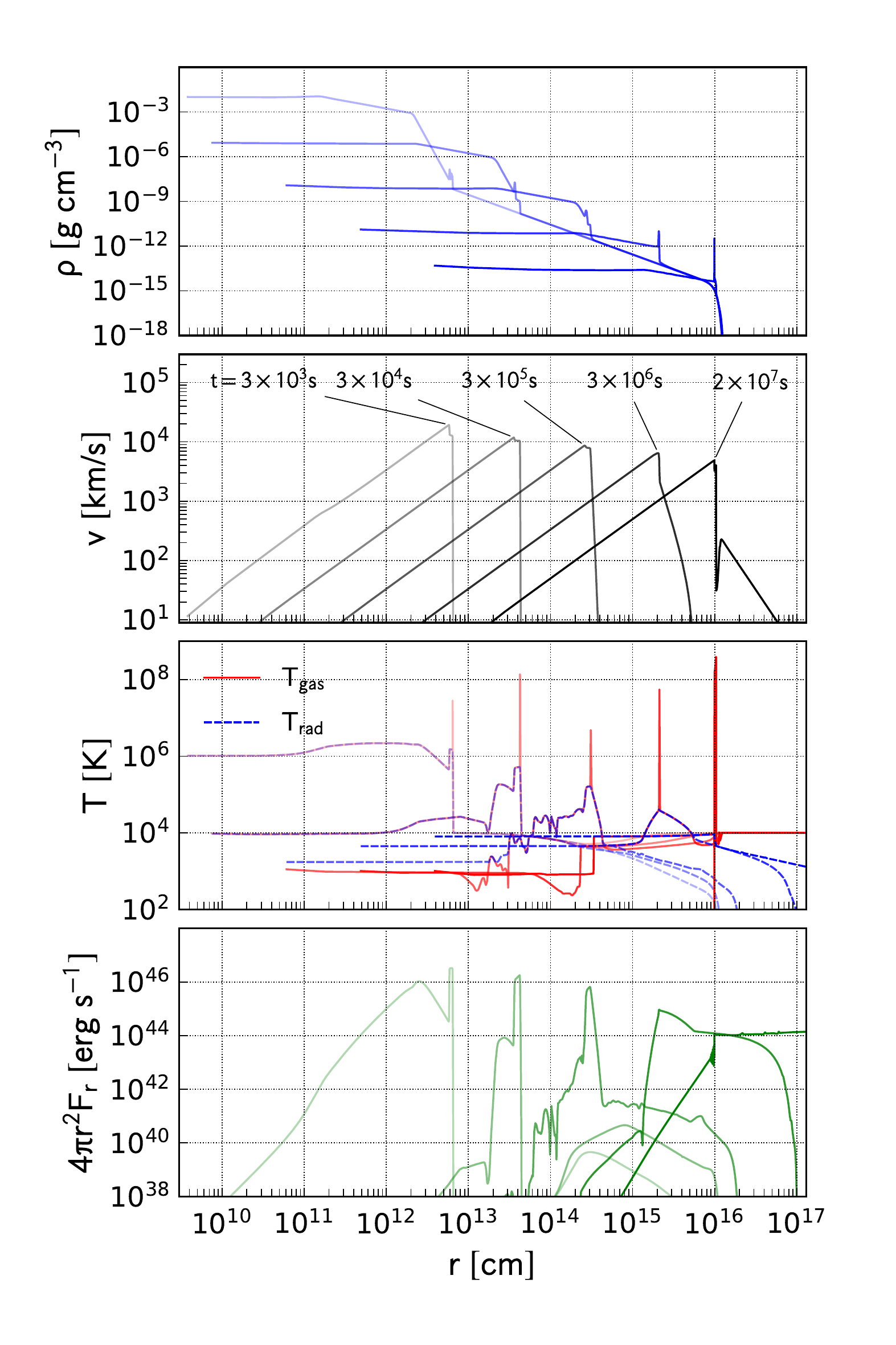}
\cprotect\caption{Radial profiles of density, velocity, gas and radiation temperature, and outgoing luminosity from top to bottom. The best-fit model with $E_\mathrm{sn}=10^{52}$ erg, $M_\mathrm{ej}=30\ M_\odot$, $M_\mathrm{csm}=8\ M_\odot$, and $R_\mathrm{csm}=10^{16}$cm is shown. 
In each panel, the radial profiles at $t=3\times 10^{3}$, $3\times 10^4$, $3\times 10^5$, $3\times 10^6$, and $2\times 10^7$ s are plotted. }
\label{fig:radial}
\end{center}
\end{figure}

\begin{figure*}
\begin{center}
\includegraphics[scale=0.6]{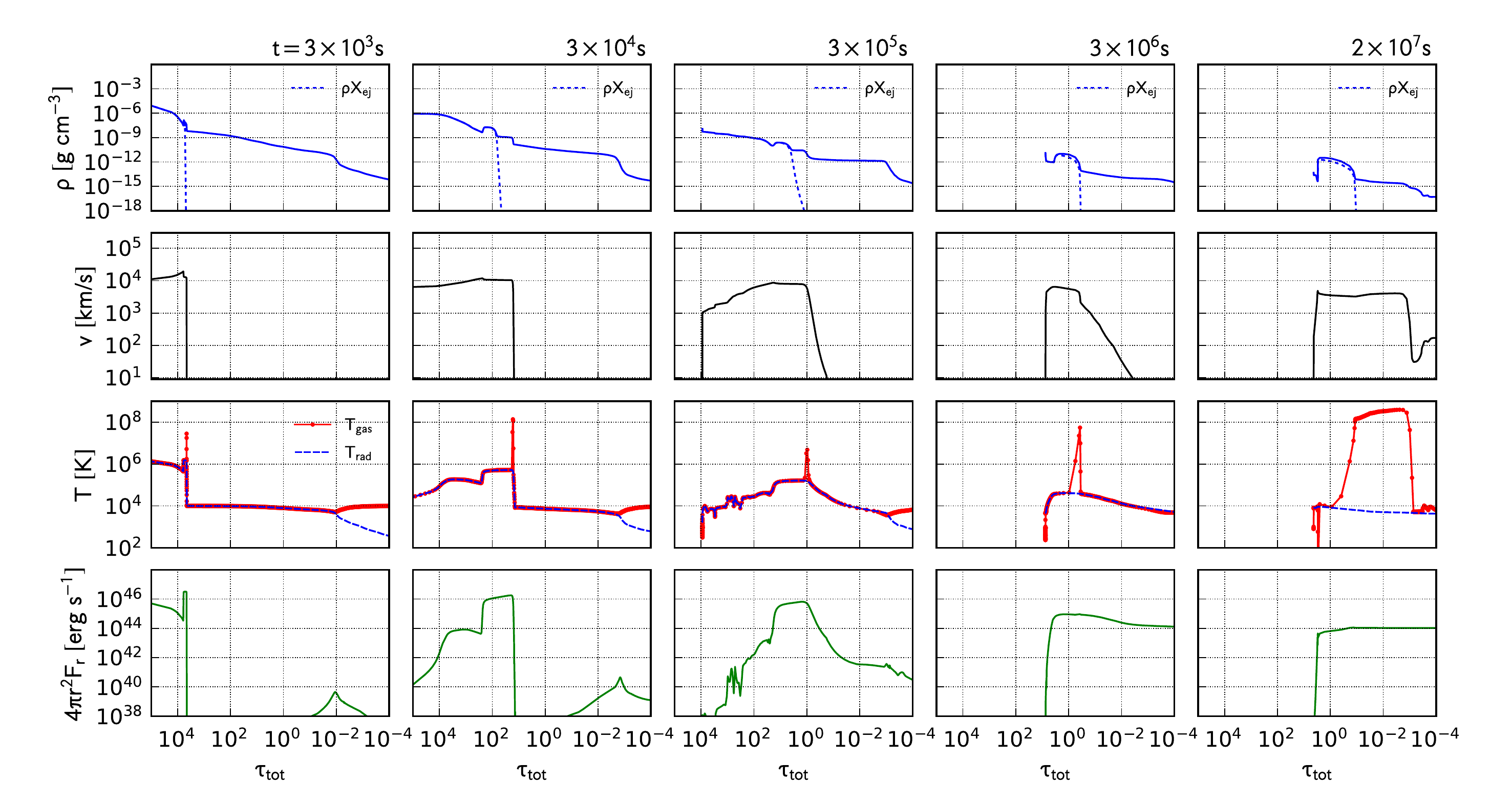}
\cprotect\caption{Physical variables as a function of the total optical depth $\tau_\mathrm{tot}$. 
The density, the velocity, the gas and radiation temperatures, and the outgoing luminosity are plotted from top to bottom)
The profiles at the same epochs as Figure \ref{fig:radial} are plotted from left to right. 
In the top panels, the dotted lines indicate the ejecta density ($\rho X_\mathrm{ej}$) distinguished from the ambient gas. 
In the gas temperature profile (red solid line), small circles indicate the locations of the numerical cells. 
}
\label{fig:profile_tau}
\end{center}
\end{figure*}

Figure \ref{fig:radial} shows the radial profiles of some physical quantities for the best-fit model. 
In this model, the shock breakout happens at several $10^6$ s and then the interaction-powered emission with the outgoing luminosity of the order of $10^{44}$ erg s$^{-1}$ starts leaking into the surrounding space as seen in the bottom panel. 
The forward shock reaches the outer edge of the CSM at $t\simeq 2\times 10^7$ s, after which the energy production rate due to the shock dissipation rapidly declines. 
In Figure \ref{fig:profile_tau}, we plot the same physical variables as Figure \ref{fig:radial}, but as a function of the total optical depth measured from the outer boundary;
\begin{equation}
    \tau_\mathrm{tot}(r)=
    \int_r^{R_\mathrm{out}}\rho(\kappa_\mathrm{a}+\kappa_\mathrm{s})dr.
\end{equation}
The structures of the shock and the precursor are better resolved in Figure \ref{fig:profile_tau}.

Since the CSM mass is a considerable fraction of the ejecta mass, the outer part of the ejecta efficiently decelerates. 
The shocked ejecta and CSM are piled up behind the forward shock, forming the so-called cold dense shell \citep[e.g.,][]{1982ApJ...258..790C,2001MNRAS.326.1448C,2004MNRAS.352.1213C,2017hsn..book..843B}. 
The maximum velocity of the ejecta initially exceeds $10^4$ km s$^{-1}$ and then slows down to $\sim 5\times 10^3$ km s$^{-1}$ at the shock emergence ($t\simeq 2\times 10^{7}$ s). 
The H$\alpha$ emission line profile of SN 2016aps exhibits a velocity full-width at half-maximum of $4090\pm 230$ km s$^{-1}$ at $80$--$350$ days after the discovery \citep{2020NatAs.tmp...78N}. 
If the observed line width represents the expansion velocity, it is consistent with the dynamical property of the best-fit model, although the line broadening may be due to the electron scattering within the CSM.

\subsection{Multi-band light curve}
\begin{figure}
\begin{center}
\includegraphics[scale=0.55]{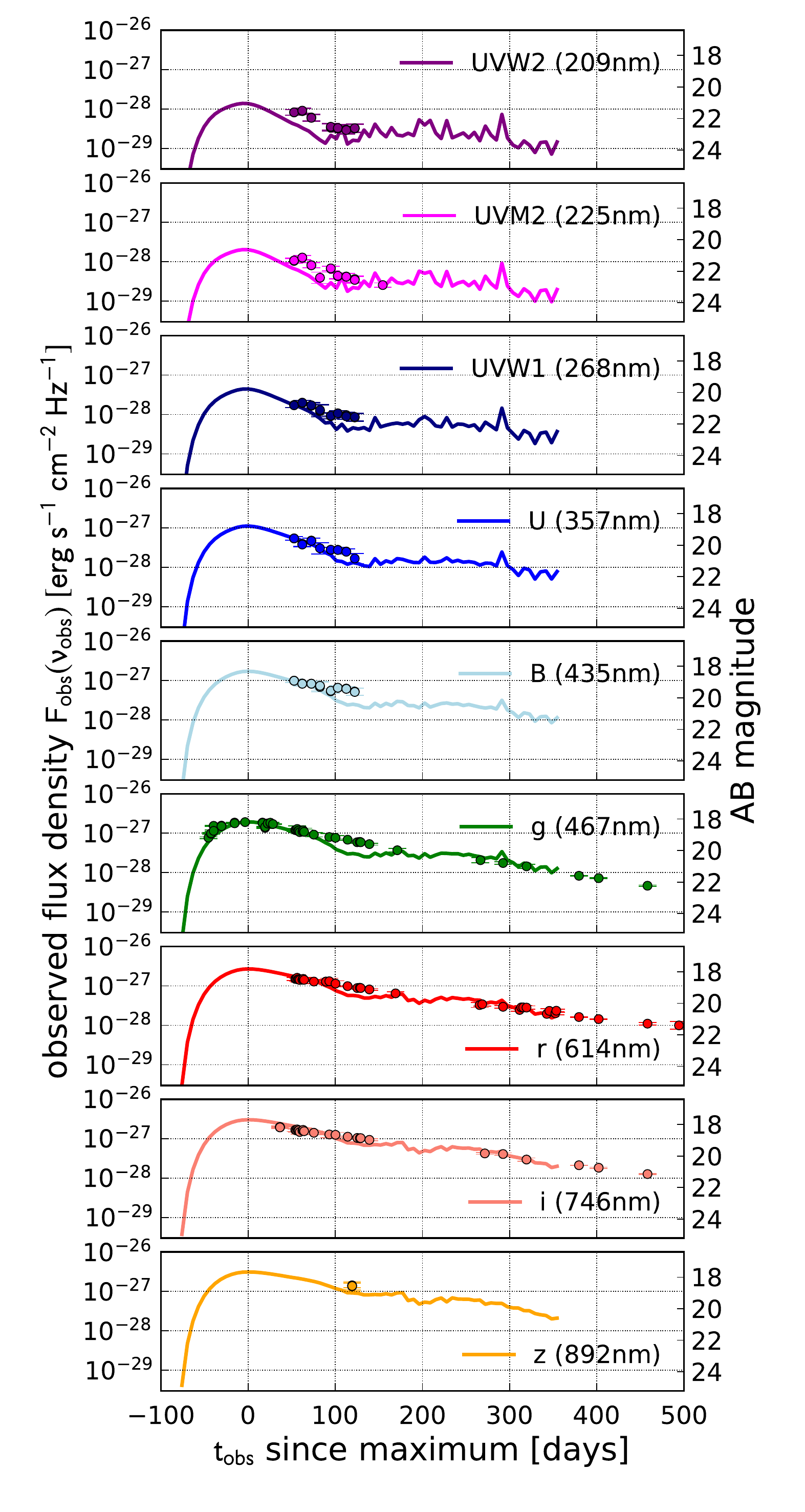}
\cprotect\caption{Multi-band light curve of the best-fit model (solid line) compared with observations of SN 2016aps (circles).}
\label{fig:multi_band_lc}
\end{center}
\end{figure}

We calculate the multi-band light curve of the best-fit model by the method described in Section \ref{sec:post_process}. 
In Figure \ref{fig:multi_band_lc}, we compare the model light curves with the observations of SN 2016aps, which are taken from the Open Supernova Catalog\footnote{https://sne.space} \citep{2017ApJ...835...64G} and mainly cover UV and optical wavelengths. 
In the upper five panels, we plotted {\it Swift}/UVOT $UVW2$-, $UVM2$-, $UVW1$-, $U$-, $B$-band light curves (the $V$-band light curve is omitted because its central wavelength is similar to the $g$- and $r$-bands), while the lower four panels present $g$-, $r$-, $i$-, and $z$-band light curves obtained by ground-based observations. 
Since the UVOT photometry is calibrated to the Vega magnitude, we convert the observed magnitudes to the AB magnitude by using the conversion factors provided by \cite{2011AIPC.1358..373B}. 
We also note that the observed magnitudes are corrected for the Galactic extinction with $E(B-V)=0.0263$ \citep{2020NatAs.tmp...78N} by using the extinction law of \cite{1989ApJ...345..245C} with $R_V=3.1$. 
The model light curve is truncated at $\sim 370$ days after the maximum because the whole ejecta becomes effectively thin, $\tau_\mathrm{eff}<1$, at the epoch and thus the effective photosphere and the color temperature $T_\mathrm{c}$ can no longer be determined. 

We find that the model light curves well agree with the observed light curves. 
This overall agreement reassures that the CSM-powered emission with almost thermal spectra gives a plausible explanation for this extremely bright SN, although slight differences are recognized. 
In particular, the model light curve exhibits a dip at $\sim 100$ days, while the observed $g$-, $r$-, $i$-band light curves almost linearly decline at the epoch. 
At this epoch, corresponding to $t=2\times 10^7$ s in the simulation, the forward shock emerges from the outer edge of the CSM (see Figure \ref{fig:radial}). 
The emergence of the forward shock is accompanied by the accelerated expansion of the shocked CSM, which makes the shocked gas cool in an adiabatic way. 
The reduced gas temperature seems to produce the dip in the model light curve. 
We note that the effect of the temperature reduction is less significant at longer wavelengths, e.g., $i$- and $z$-bands, probably because they are in the Rayleigh-Jeans part of the almost blackbody spectra. 
This effect may be avoided by exploring a more appropriate outer cut-off in the CSM density at $r=R_\mathrm{csm}$ (Equation \ref{eq:rho_csm}), which is both theoretically and observationally uncertain. 
Nevertheless, such a slight modification on the interface between the massive CSM and the normal stellar wind would not change the overall picture, i.e., several $10\ M_\odot$ SN ejecta colliding with $\sim 10\ M_\odot$ CSM with $R_\mathrm{csm}\simeq 10^{16}$ cm as the best explanation for SN 2016aps. 

\begin{figure}
\begin{center}
\includegraphics[scale=0.55]{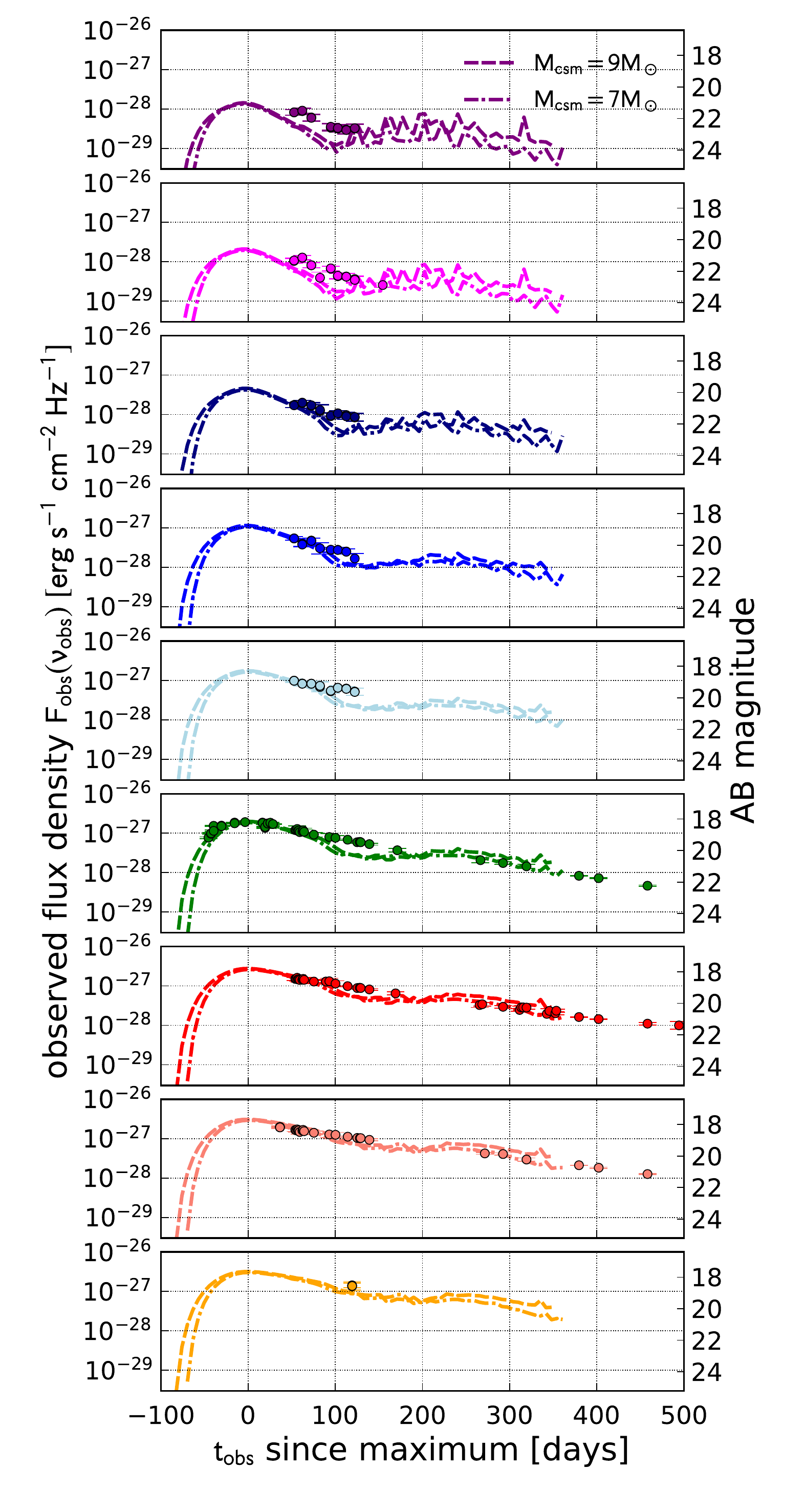}
\cprotect\caption{Same as Figure \ref{fig:multi_band_lc}, but for the models with $M_\mathrm{csm}=7$ (dash-dotted) and $9\ M_\odot$ (dashed) in the same model series \verb|M30R10E10|.}
\label{fig:multi_lc_comparison}
\end{center}
\end{figure}

We plot the models in the same model series as the best-fit model but with slightly different CSM  masses, $M_\mathrm{csm}=7$ and $9\ M_\odot$, in Figure \ref{fig:multi_lc_comparison}. 
This comparison demonstrates how the model light curves depend on the CSM mass and how large uncertainty is associated with the light curve fitting. 
As seen in the $g$-band light curves with the early data available, the model with the smaller $M_\mathrm{csm}=7\ M_\odot$ rises a bit more sharply than the best-fit model, while the model with the larger $M_\mathrm{csm}=9\ M_\odot$ exhibits a slower rise. 
Although the model with $M_\mathrm{csm}=8\ M_\odot$ gives the best-fit light curve, models with slightly different CSM masses by $\simeq 1\ M_\odot$ appear to fit the observations well.

Finally, we make a remark on the spectral evolution. 
In our best fit model, the effective optical depth of the whole ejecta decreases to unity around $\sim 370$ days, after which the ejecta is supposed to enter the nebular phase. 
\cite{2020NatAs.tmp...78N} provide the spectra of SN 2016aps at several 100 days after the maximum. 
The blue continuum around $4000$--$5000\ \mathrm{\AA}$ remains in the spectrum at $+139$ days. 
The residual continuum remains at $+345$ days, while it becomes rather flat and instead Balmer emission lines significantly contribute to the total emission at $+536$ days.  
Since the transition from the photospheric to the nebular spectrum usually happens in a gradual manner and the optical thickness of the ejecta also depends on the wavelength, it is not easy to determine when exactly the whole ejecta becomes transparent. 
Nevertheless, the dynamical evolution of the ejecta looks roughly consistent with the observed spectral evolution. 
In our simulations based on gray radiative transfer, we do not treat wavelength-dependent opacity (line opacities, in particular) properly, although we roughly estimate their impact on the late-time light curve in Figure \ref{fig:opacity_effect}. 
Since the observed spectra show prominent line emission, the contribution of line opacities would become significant at later epochs and then the reddening caused by line blanketing may modify the multi-band light curve. 
Multi-group radiation-hydrodynamic simulations combined with spectral synthesis calculations are ultimately required for a fair comparison between numerical models and the observations.


\section{Discussion}\label{sec:discussion}
\subsection{Model parameters for SN 2016aps}
The light curve fitting in Section \ref{sec:results} suggests that SN 2016aps is most likely explained by an energetic SN ejecta with $M_\mathrm{ej}=30\ M_\odot$ and $E_\mathrm{sn}=10^{52}$ erg colliding with a wind-like CSM with $M_\mathrm{csm}\simeq 8$ $M_\odot$ and $R_\mathrm{csm}\simeq 10^{16}$ cm. 
The model parameters are roughly consistent with the analytic consideration in Section \ref{sec:parameter_estimate}. 
With the estimated total mass of the ejecta and the CSM, and the stellar-mass compact remnant included, the initial mass of the exploding star should be at least $\gtrsim 40\ M_\odot$. 
Given that the SN explosion was highly energetic with the explosion energy of the order of $10^{52}$ erg, i.e., hypernovae, a massive star with the initial mass larger than $40\ M_\odot$ and ejecting a part of the hydrogen envelope shortly before the core-collapse is a likely progenitor. 

The suggested progenitor is different from the scenario originally proposed by \cite{2020NatAs.tmp...78N}, who suggest the collision of much more massive SN ejecta and CSM ($E_\mathrm{sn}>3\times 10^{52}$ erg, $M_\mathrm{ej}=50$--$180\ M_\odot$, and $M_\mathrm{csm}=40$--$150\ M_\odot$) as a likely explanation of SN 2016aps. 
Their argument was interesting because the total mass of the ejecta and CSM is within the mass range of the pair-instability SNe ($140$--$260 \ M_\odot$; \citealt{2002ApJ...567..532H,2002ApJ...565..385U}). 
The disagreement between our model and \cite{2020NatAs.tmp...78N} could partly be explained by the different treatment of radiation transport in the light curve models. 
While we carried out 1D radiation-hydrodynamic simulations with the so-called two-temperature approximation, \verb|MOSFiT| is based on a single-zone model for the expanding ejecta with the heating source deeply embedded \citep{1982ApJ...253..785A}. 
The radiation-hydrodynamic simulations spatially resolve the interaction layer between the ejecta and the CSM, which moves outward with time, and solves the radiation transport in the CSM in a more sophisticated way. 
\cite{2015MNRAS.449.4304D} have conducted 1D multi-group radiation-hydrodynamic simulations of superluminous SNe-IIn. 
Their models with the initial kinetic energy of $10^{52}$ erg predict bolometric peak luminosities and evolutionary timescales similar to our models, although the ejecta and CSM masses of their model ($M_\mathrm{ej}=9.8\ M_\odot$ and $M_\mathrm{csm}=17.3\ M_\odot$) are different from ours by a factor of a few. 
This agreement among numerical simulations suggests that the model parameters constrained by radiation-hydrodynamic simulations might be systematically different from those obtained by more simplified single-zone models and analytic scaling relations.

\cite{2012ApJ...746..121C} and \cite{2013MNRAS.428.1020M} tried the light curve modeling of the superluminous SN-IIn 2006gy in semi-analytic and numerical ways, respectively. 
Their best-fit models also exhibit differences in some model parameters, such as the ejecta mass. 
Later, \cite{2013ApJ...773...76C} compared the semi-analytic and numerical models in more detail and pointed out a difficulty of a simple maximum likelihood parameter estimate for interaction-powered SNe. 
While semi-analytic models certainly provide a convenient way to search for appropriate parameters in a large parameter space, different sets of parameters sometimes provide equally well fitting results. 
A thorough investigation on the possible systematic differences between semi-analytic and numerical light curve models for interaction-powered SNe would be beneficial, but is beyond the scope of this paper. 

We also note that the nickel-powered scenario is unlikely to explain the emission properties of SN 2016aps. 
As we have shown in the middle panel of Figure \ref{fig:lc_comparison}, the energy deposition by $^{56}$Co decay matches the declining part of the light curve only when an extremely large amount of nickel mass, $\sim 70\ M_\odot$ is assumed. 
Such a huge amount of radioactive nickel is only produced in PISNe around the massive end ($\sim 300\ M_\odot$). 
In addition, the decline rate of the late-time light curve similar to $^{56}$Co decay rate implies almost full-trapping of nuclear gamma-rays and positrons produced by $^{56}$Co decay and therefore requires ejecta mass even larger than the nickel mass. 
Considering the fact that the nickel masses estimated for broad-lined Type-Ic SNe are at most $\sim 1\ M_\odot$ \citep{2017AdAst2017E...5C}, radioactive nickel (if any) contributes to the emission from SN 2016aps only in a minor way.

To summarize, we suggest that the collision between massive SN ejecta with a relatively massive hydrogen-rich CSM is a viable scenario explaining SN 2016aps, but the progenitor does not necessarily require the initial mass exceeding $100\ M_\odot$.

\subsection{Mass-loss episode of SN 2016aps}
Although the origin of hypernovae is still unclear, they are predominantly the explosion of a bare carbon-oxygen core with little trace of hydrogen and helium in their spectra, except for a few hydrogen-rich events,  such as OGLE-2014-SN-073 (see, e.g., \citealt{2017NatAs...1..713T} and references therein). 
Therefore, the presence of SN 2016aps implies the possibility that massive stars ending their lives as hypernovae may occasionally eject their hydrogen-rich envelope at the final moment of their evolution and produce superluminous SNe-IIn. 
The CSM mass and radius constrained by our light curve modeling suggest that an intense mass-loss episode shortly before the core-collapse is required to produce the CSM confined within $R_\mathrm{csm}\simeq 10^{16}$ cm. 

For a wind velocity $v_\mathrm{w}$, the enhanced mass-loss should have initiated at
\begin{equation}
    t_\mathrm{w}=\frac{R_\mathrm{csm}}{v_\mathrm{w}}\simeq 30\left(\frac{R_\mathrm{csm}}{10^{16}\mathrm{cm}}\right)
    \left(\frac{v_\mathrm{w}}{100\ \mathrm{km\ s}^{-1}}\right)^{-1} \mathrm{yrs},
\end{equation}
before the core-collapse. 
We have used $v_\mathrm{w}=100$ km s$^{-1}$ as our fiducial value. 
Unfortunately, early spectra of SN 2016aps do not resolve the narrow absorption/emission line component \citep{2020NatAs.tmp...78N} and therefore cannot infer the wind velocity accurately. 
The mass-loss rate averaged over the period is required to be
\begin{eqnarray}
    \dot{M}
    &=&
    0.3\ \mathrm{M_\odot\ yr}^{-1}
    \left(\frac{M_\mathrm{csm}}{10\ M_\odot}\right)
    \nonumber\\
    &&\times
    \left(\frac{R_\mathrm{csm}}{10^{16}\mathrm{cm}}\right)^{-1}
    \left(\frac{v_\mathrm{w}}{100\ \mathrm{km\ s}^{-1}}\right).
\label{eq:Mdot}
\end{eqnarray}
This mass-loss rate is much higher than the typical mass-loss rate of Galactic red supergiants or Wolf-Rayet stars, but is consistent with those required for luminous SNe-IIn \citep[e.g.,][]{2012ApJ...744...10K,2013A&A...555A..10T,2014ARA&A..52..487S}. 
Therefore, the final mass-loss activity for SN 2016aps may be driven by the same mechanism as other SNe-IIn. 
The mass ejection mechanism to realize SNe-IIn is also unclear and thus extensively debated. 
The energy deposition onto the stellar envelope by stellar activities shortly before the core-collapse \citep{2010MNRAS.405.2113D,2012MNRAS.423L..92Q,2014ApJ...780...96S,2017MNRAS.470.1642F,2018MNRAS.476.1853F,2019ApJ...877...92O,2019MNRAS.485..988O,2020A&A...635A.127K} or binary effects, such as the common envelope mass ejection \citep{2012ApJ...752L...2C,2013ApJ...764L...6S,2020ApJ...892...13S}, have been proposed as the responsible mechanism.

It is worth comparing the estimated parameters with those of other SNe-IIn. 
SN 2010jl \citep[e.g.,][]{2011ApJ...732...63S,2012AJ....144..131Z,2014ApJ...797..118F,2014ApJ...781...42O} is among the best studied SN-IIn. 
\cite{2014ApJ...797..118F} provide a comprehensive observational study of SN 2010jl up to $\sim$ 1000 days after the first detection. 
They estimated the mass-loss rate and the CSM mass to be $\dot{M}\sim 0.1\ M_\odot$ yr$^{-1}$ and $\gtrsim 3\ M_\odot$. 
\cite{2015MNRAS.449.4304D} performed a radiation-hydrodynamic simulation of a $\sim 10\ M_\odot$ SN ejecta with $10^{51}$ erg colliding with $\sim 3\ M_\odot$ CSM and found a good agreement with their model light curve and the bolometric light curve of SN 2010jl. 
The mass-loss rate estimated in Equation \ref{eq:Mdot} and the best-fit CSM mass of $8\ M_\odot$ for SN 2016aps agrees with those of SN 2010jl within a factor of a few. 
Instead, SN 2016aps is much more luminous than SN 2010jl, which showed the peak bolometric luminosity of $3\times 10^{43}$ erg s$^{-1}$. 
The similar mass-loss rate and the CSM mass suggest that the progenitor of SN 2016aps has experienced a mass-loss episode that commonly happens in other typical SNe-IIn. 
It would be the much larger explosion energy that makes SN 2016aps extremely bright with the peak luminosity 10 times higher than SN 2010jl. 

SN 2006gy is another well-studied superluminous SN-IIn \citep{2007ApJ...659L..13O,2007ApJ...666.1116S}. 
Early studies interpreted this object as a core-collapse event, although it is recently proposed that a Type Ia SN embedded in a massive CSM better explains the observed properties \citep{2020Sci...367..415J}. 
For example, \cite{2013MNRAS.428.1020M} performed a series of radiation-hydrodynamic simulations of interaction-powered SNe. 
They found that the multi-band light curve of SN 2006gy is well explained by an energetic explosion with $>4\times10^{51}$ erg and a $\sim 15\ M_\odot$ CSM. 
When adopting the interaction-powered CCSN scenario, the estimated average mass-loss rate of $\sim 0.1\ M_\odot$ yr$^{-1}$ is again similar to our estimate for SN 2016aps, while SN 2006gy requires a bit more massive CSM. 
The comparison of SN 2016aps with the prototypical SN-IIn 2010jl and the superlumninous SN-IIn 2006gy indicates that the explosion energy of the embedded SN is a primary factor determining the brightness rather than the mass-loss rate.

\subsection{Pulsational pair-instability scenario}
Although our ejecta and CSM mass estimate does not necessarily require pair-instability SNe with $>100\ M_\odot$, massive stars that experience the pulsational pair-instability can be a candidate progenitor system \citep{2007Natur.450..390W,2015A&A...573A..18M,2017ApJ...836..244W}. 
Because SNe with explosion energies of the order of $10^{52}$ erg are predominantly hydrogen-poor (in particular, broad-lined Type-Ic SNe), it is reasonable to assume that the ejecta responsible for SN 2016aps is a helium or carbon-oxygen core. 
A helium core more massive than $30$--$40\ M_\odot$ with a sub-solar metallicity is expected to experience intensive mass-loss due to the pair-instability \citep[see,][for a recent study]{2017ApJ...836..244W}. 
Assuming a total core mass for SN 2016aps corresponding to the ejecta mass of $M_\mathrm{ej}=30\ M_\odot$ plus a remnant compact object (a $\sim 5$--$10\ M_\odot$ black hole to power the hypernova), our best-fit model satisfies the pulsational pair-instability condition. 
In addition, SN 2016aps occurred in a sub-solar metallicity environment ($Z\sim 0.4Z_\odot$; \citealt{2020NatAs.tmp...78N}). 
These agreements suggest the pulsational pair-instability as an intriguing and promising progenitor channel for SN 2016aps. 

\cite{2015A&A...573A..18M} studied pulsational behaviors of PISN progenitors in their red sugergiant stage and claimed that pulsation-induced mass-loss could explain the high mass-loss rate inferred from superluminous SNe-IIn observations.  
\cite{2017ApJ...836..244W} performed a comprehensive study on pulsational pair-instability SNe and presented some models that show pre-supernova mass of several 10 $M_\odot$ after having ejected their hydrogen-rich envelope of the order of $10\ M_\odot$ several years before the core-collapse. 
He also pointed out the possibility that a hypernova explosion may occur in such a situation and produce a bright SNe with the peak luminosity close to $10^{45}$ erg s$^{-1}$. 
Our best fit model is broadly in agreement with the scenario of pulsational pair-instability followed by a hypernova explosion.

\subsection{Hypernova explosions in massive CSM}
SN 2016aps is an unambiguous example of SNe-IIn with the radiated energy well exceeding $10^{51}$ erg. 
SN 2016aps-like events harboring hypernova explosions in their massive CSMs are expected in the on-going and future transient surveys, such as Zwicky Transient Facility \citep{2019PASP..131a8002B} and the Vera C. Rubin observatory\footnote{https://www.lsst.org}. 
\cite{2020NatAs.tmp...78N} found that an SN 2016aps-like event could be detected up to $z\sim 2$ with the Rubin observatory, and even higher redshift $z=5$ with the James Webb Space Telescope\footnote{https://www.jwst.nasa.gov}.
This opens up the possibility that the final activities of very massive stars with the initial masses of several $10\ M_\odot$ could be probed in the high-$z$ universe. 
Although the rate of SN 2016aps-like events is uncertain, requiring the initial mass of $>40\ M_\odot$ instead of the originally proposed condition, $100$--$300\ M_\odot$, may relax the constraint on the rareness of such events for the standard initial mass function. 
The rate of SN 2016aps-like events is also important for understanding the population of luminous SNe-IIn. 
The rate of SN 2016aps-like event together with some statistical samples of SNe-IIn \citep[e.g.,][]{2014ApJ...788..154O,2020A&A...637A..73N} constrains how the luminosity function of SNe-IIn extends to the highest luminosity, which would give us some hint on the evolutionary scenario or the mechanism to produce massive CSMs.

In this work, we provide the $L_\mathrm{bol,peak}$--$t_\mathrm{rise}$ and $E_\mathrm{rad}$--$t_\mathrm{rise}$ relations in Figure \ref{fig:Lpeak_Tpeak} and used them to narrow down the parameter space. 
The same exercise can be applied for possible other superluminous SNe-IIn with the radiated energy exceeding $10^{51}$ erg. 
For example, an SN 2016aps-like event with the peak bolometric luminosity as high as $10^{45}$ erg s$^{-1}$ indicates a highly energetic explosions with a compact CSM. 
The comparison between a statistical sample of such superluminous SNe-IIn and the theoretical predictions may eventually unveil the origin of CSM and the evolution of very massive stars toward the core-collapse.

\section{Summary}\label{sec:summary}
In this work, we performed the light curve modeling of the superluminous SN-IIn 2016aps. 
Assuming that the emission is predominantly powered by the collision of a highly energetic SN ejecta with a wind-like CSM, we constrain the appropriate model parameters for SN 2016aps. 
As a result, we find that the combination of an SN ejecta with the mass $M_\mathrm{ej}=30\ M_\odot$ and $E_\mathrm{sn}=10^{52}$ erg and a CSM with the mass $M_\mathrm{csm}=7$--$9\ M_\odot$ and the radius $R_\mathrm{csm}=10^{16}$ cm most likely explains the multi-band light curve of SN 2016aps. 
This finding suggests that very massive stars potentially producing hypernova explosions can be superluminous SNe-IIn. 
However, the reason why SN 2016aps had such a massive hydrogen-rich CSM in its vicinity while hypernovae are almost exclusively Type-Ic SNe is unclear. 
This difference might be ascribed to the different mass-loss mechanisms, normal stellar wind for Type-Ic SNe and eruptive mass-loss for SN 2016aps, the latter of which may be realized in a low metallicity environment. 
Although SN 2016aps is still an example of hypernova explosions with massive CSMs, future surveys and detection of similar events along with light curve modelings would clarify how special SN 2016aps-like events are and their nature.

\acknowledgements
A.S. acknowledges support by Japan Society for the Promotion of Science (JSPS) KAKENHI Grand Number JP19K14770. M.N.~is supported by a Royal Astronomical Society Research Fellowship. 
Numerical simulations were carried out by Cray XC50 system operated by Center for Computational Astrophysics, National Astronomical Observatory of Japan. 

\software{Matplotlib (v3.2.1; \citealt{2007CSE.....9...90H})
}

\appendix
\section{Post-process calculations}\label{sec:post_process}

In the following steps, we describe the numerical procedures of the post-process calculations. 
We denote physical variables in the comoving frame of emitting material by letters with overbars. 
For example, the comoving frequency is denoted by $\bar\nu$, while $\nu$ represents the frequency in the laboratory frame. 
For a photon with the direction cosine $\mu$, the frequencies in these two different rest frames are related with each other by Lorentz transformation,
\begin{equation}
    \frac{\bar\nu}{\nu}=\Gamma(1-\beta\mu),
\end{equation}
where $\beta$ is the radial velocity in the unit of $c$ and  $\Gamma=(1-\beta^2)^{-1/2}$ is the corresponding Lorentz factor. 

\begin{figure}
\begin{center}
\includegraphics[scale=0.3]{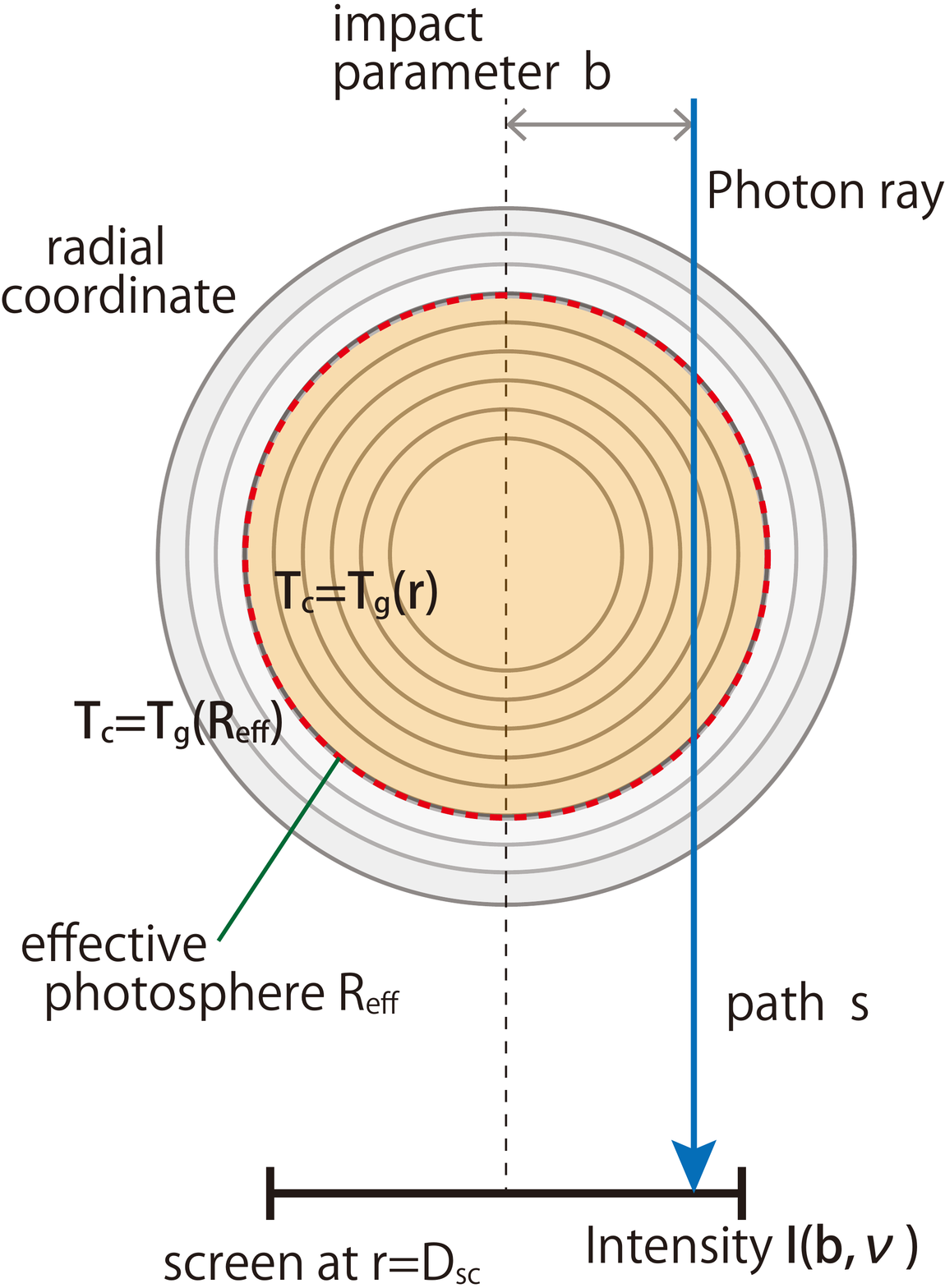}
\cprotect\caption{Schematic representation of the post-process ray-tracing calculation. 
The transfer equation is integrated along a path $s$ on the photon ray. 
}
\label{fig:ray_tracing}
\end{center}
\end{figure}

We consider $N_b(=1024)$ parallel rays emanating the spherical ejecta with different impact parameters $b$. 
Figure \ref{fig:ray_tracing} schematically represents the ray-tracing calculations. 
We conduct ray-tracing calculations on each ray for different frequencies $\nu$ and obtain the intensity map $I(b,\nu)$ on the screen normal to the radial direction at a distance $D_\mathrm{sc}=1.5\times 10^{17}$ cm. 
The frequency range is given by $\nu\in[10^{12}\mathrm{Hz},10^{17}\mathrm{Hz}]$ and is discretized into $N_\nu(=128)$ logarithmically spaced points. 
For a given intensity map $I(b,\nu)$ on the screen, the luminosity per unit frequency $L(\nu)$ in the source rest-frame is calculated as follows,
\begin{equation}
    L(\nu)=4\pi \int I(b,\nu) dS=8\pi^2\int I(b,\nu)bdb,
\end{equation}
where $dS=2\pi bdb$ is the area element of the annulus corresponding to the impact parameter $b$. 
The area of the integration is taken to be sufficiently large to cover the optically thick part of the ejecta. 

\subsection{Ray-tracing along a ray}
The ray-tracing calculation on a single ray for a frequency $\nu$ is carried out in the following way. 
We integrate the transfer equation for the intensity $I(\nu)$ in the laboratory frame along a path $s$,
\begin{eqnarray}
    \frac{dI(\nu)}{ds}&=&
    \left(\frac{\nu}{\bar\nu}\right)^2
    \left[\bar{\kappa}_\mathrm{a}(\bar\nu)+\bar{\kappa}_\mathrm{s}(\bar\nu)
    \right]\bar{S}(\bar\nu)
    \nonumber \\&&
    -\left(\frac{\bar\nu}{\nu}\right)
    \left[\bar{\kappa}_\mathrm{a}(\bar\nu)+\bar{\kappa}_\mathrm{s}(\bar\nu)
    \right]I(\nu),
\end{eqnarray}
where $\bar{\kappa}_\mathrm{a}(\bar\nu)$ and $\bar{\kappa}_\mathrm{s}(\bar\nu)$ are the frequency-dependent absorption and scattering coefficients and $\bar{S}(\bar\nu)$ is the source function. 
The integration of the transfer equation is straightforward once the source function is known. 
We adopt the following source function,
\begin{equation}
    \bar{S}(\bar\nu)=
    \frac{\bar{\kappa}_\mathrm{a}(\bar\nu)B(\bar\nu,T_\mathrm{g})+\bar{\kappa}_\mathrm{s}(\bar\nu)\bar{J}(\bar\nu)}{\bar{\kappa}_\mathrm{a}(\bar\nu)+\bar{\kappa}_\mathrm{s}(\bar\nu)},
\end{equation}
\citep[e.g.,][]{1979rpa..book.....R}, where absorbed photons are re-emitted according to the Planck function with the local gas temperature $T_\mathrm{g}$, while scattered photons are redistributed according to the comoving mean intensity $\bar{J}(\bar\nu)$. 

Since we assume electron scattering, the scattering opacity is frequency-independent and therefore equal to the gray scattering coefficient in Equation \ref{eq:kappa_es}, $\bar\kappa_\mathrm{s}(\nu)=\kappa_\mathrm{s}$. 
We adopt the absorption coefficient whose Planck mean is equal to the gray absorption coefficient, Equation \ref{eq:kappa_a},
\begin{eqnarray}
    \bar{\kappa}_\mathrm{a}(\bar\nu)&=&
    2.2\times 10^{-36}\chi_\mathrm{ion}
    (1+X_\mathrm{h})(X_\mathrm{h}+X_\mathrm{he})\rho T_\mathrm{g}^{-1/2}
    \nonumber\\
    &&\times
    \nu^{-3}\left[1-\exp\left(-\frac{h\nu}{k_\mathrm{B}T_\mathrm{g}}\right)\right]\ \mathrm{cm}^2\ \mathrm{g}^{-1}
    ,
\end{eqnarray}
so that the ray-tracing results are consistent with the corresponding simulation results based on the gray opacities.

The mean intensity is usually obtained from the solution $I(\nu)$ of the transfer equation, which makes the problem complicated. 
In this work, we adopt the following approximated mean intensity instead of fully solving the transfer equation. 
From the numerical simulations, we obtain the distributions of the radiation energy density $E_\mathrm{r}$ and flux $F_\mathrm{r}$. 
Therefore, we can estimate the frequency-integrated comoving radiation energy density as follows,
\begin{equation}
\bar{E}_\mathrm{r}\simeq E_\mathrm{r}-2\beta F_\mathrm{r},
\end{equation}
by neglecting higher-order terms $\propto {\cal O}(\beta^2)$. 
We further assume that the mean intensity is proportional to the Planck function with a color temperature $T_\mathrm{c}$,
\begin{equation}
    \bar{J}(\bar\nu)=\frac{\bar{E}_\mathrm{r}}{a_\mathrm{r}T_\mathrm{c}^4}B(\bar\nu,T_\mathrm{c}),
\end{equation}
where $a_\mathrm{r}$ is the radiation constant. 
In this approximation, the transfer equation can be integrated in a straightforward way for a given color temperature distribution. 

\subsection{Photosphere and color temperature}
The color temperature is determined in the following way. 
First, we determine the effective photosphere $r=R_\mathrm{eff}$ at which the effective optical thickness $\tau_\mathrm{eff}(r)$ along the radial direction from $r$ to $R_\mathrm{out}$,
\begin{equation}
    \tau_\mathrm{eff}(r)=\int_r^{R_\mathrm{out}} \rho(r)\sqrt{\kappa_\mathrm{a}(\kappa_\mathrm{s}+\kappa_\mathrm{a})}dr,
\end{equation}
\citep[e.g.,][]{1979rpa..book.....R} with gray opacities, is equal to unity. 
The local gas temperature at the effective photosphere well represents the color temperature of the emission escaping into the surrounding space. 
On the other hand, the gas inside the effective photosphere is well coupled with radiation through absorption and emission and thus the color temperature is equal to the local gas temperature. 
Therefore, we assume that the color temperature at radius $r$ is given by
\begin{equation}
    T_\mathrm{c}(r)=\left\{
    \begin{array}{cl}
    T_\mathrm{g}(r)&\ \ \ \mathrm{for}\ r\leq R_\mathrm{eff},\\
    T_\mathrm{g}(R_\mathrm{eff})&\ \ \ \mathrm{for}\ R_\mathrm{eff}<r.
    \end{array}
    \right.
\end{equation}

\subsection{Observed flux density}
Finally, we obtain the observed flux density for a source at the redshift $z$ by taking into account the following K-correction,
\begin{equation}
    F_\mathrm{obs}(\nu_\mathrm{obs})=\frac{(1+z)L(\nu)}{4\pi D_\mathrm{L}^2},
\end{equation}
where the observed-frame frequency is given by $\nu_\mathrm{obs}=\nu/(1+z)$ and $D_\mathrm{L}$ is the luminosity distance ($D_\mathrm{L}=1.4$ Gpc at $z=0.2657$). 
We also take into account the following redshift correction for the observed time,
\begin{equation}
    t_\mathrm{obs}=(1+z)t,
\end{equation}
when comparing the model multi-band light curves with observed ones.


\bibliography{refs}{}
\bibliographystyle{aasjournal}



\end{document}